\documentclass[aps,prd,reprint,nofootinbib,showpacs,longbibliography]{revtex4-2}
\usepackage{amssymb}
\usepackage{amsmath}
\usepackage{amsfonts}
\usepackage{amsthm}
\usepackage{graphicx}
\usepackage{color}
\usepackage[hang,nooneline]{subfigure}

\begin{document}

\title{Potentials for general-relativistic geodesy}

\author{Claus L{\"a}mmerzahl${}^{1,2,3}$ and Volker Perlick${}^1$}

\affiliation{${}^1$ Center for Applied Space Technology and Microgravity (ZARM), University of Bremen, 28359 Bremen, Germany \\ 
${}^2$ Institute of Physics, Carl von Ossietzky University Oldenburg, 26111 Oldenburg, Germany \\ 
${}^3$ Gauss-Olbers Space Technology Transfer Centre, c/o University of Bremen, Am Fallturm, 28359 Bremen, Germany}

\begin{abstract}
Geodesy in a Newtonian framework is based on the Newtonian gravitational potential. The general-relativistic gravitational field, however, is not fully determined by a single potential. The vacuum field around a stationary source can be decomposed into two scalar potentials and a tensorial spatial metric, which together serve as the basis for general-relativistic geodesy. One of the scalar potentials is a generalization of the Newtonian potential while the second one describes the influence of the rotation of the source on the gravitational field for which no non-relativistic counterpart exists. In this paper the operational realizations of these two potentials, and also of the spatial metric, are discussed. For some analytically given spacetimes the two potentials are exemplified and their relevance for practical geodesy on Earth is outlined. 
\end{abstract}

\maketitle


\section{Introduction}

\vspace{-0.25cm}

Beside astronomy, geodesy is one of the oldest science. It is about the shape of the Earth, its orientation, and its gravitational field. While its shape, the topography, can be observed directly e.g. from space with satellites equipped with Lidar systems, its orientation is inferred e.g. from VLBI observations and from direct measurements of the rotation of the Earth, e.g.~\cite{Gebaueretal2020}, and today provided by the International Earth Rotation Service IERS \cite{IERS}. The gravitational field can be determined from the measurements of gravimeters like falling corner cubes \cite{MarsonFaller1986}, superconducting gravimeters with a resolution better than $1\;\text{nm}/\text{s}^2$ \cite{Hinderer2015} and gradiometers. These devices measure the vector of the gravitational acceleration $\boldsymbol{g}$ which is the gradient of the gravity potential $W = U + \Omega^2 r^2$ where $U$ is the Newtonian gravitational potential and $\Omega^2 r^2$ is the centrifugal potential. Equipotential surfaces, and in particular the geoid, are constructed from $\boldsymbol{g}$-measurements and a procedure called geodetic leveling \cite{Young1990} which possesses an inconsistency within Europe of approx. 1 m. 

New developments on the experimental side open up new possibilities and improved precision to measure the gravity field of the Earth: (i) atom interferometers serve as a new class of gravi- and gradiometers \cite{Manoret2018} and they are sensitive to differences of the gravity potential, (ii) clocks through the general-relativistic redshift are also sensitive to  differences of the gravity potential \cite{Mehlstaubler:2018saw}, and (iii) the new laser ranging interferometer LRI 
on board of the GRACE Follow-On satellites, which were
launched in 2018, yield improved data for determining the gravity field of the Earth on a global scale. The precision of the LRI of 1 nm \cite{Abichetal2019} and in particular the use of the gravitational redshift of clocks now make it mandatory to describe these measurements within the formalism of General Relativity. 

The main task of geodesy is to determine and to characterize the gravitational field of a compact gravitating body like the Earth. On the Newtonian level a characteristic quantity is the geoid which is a certain surface of a constant Newtonian gravity potential possessing the topology of a sphere. Also in General Relativity it was possible to define a fully general-relativistic geoid which can be determined through clocks or through gravimetric measurements \cite{Philippetal2017}. However, since the gravitational field within General Relativity possesses more degrees of freedom than in Newtonian gravity (we have 10 metrical components compared with one Newtonian gravitational potential) one may wonder whether there might also be more than one kind of geoid within the framework of Gdeneral Relativity. 

In fact, in this paper we define with the help of a second potential a second geoid which is related to the gravitational field of a stationary rigid body. While the first geoid is mainly related to the mass density of the gravitating source, the second geoid is related to the mass \textit{current} density of the source, in particular its rotation, i.e., to the gravitomagnetic part of the general-relativistic gravitational field. Both geoids are related to the stationarity of the gravitational field and the second one requires, in addition, that Einstein's vacuum field equation is satisfied, i.e., it is defined only outside of the source. The other degrees of freedom of the gravitational field are included in the remaining metrical components in a 3-dimensional rest frame. 

In developing the notions we first state the model of the Earth which we assume to rotate rigidly and to be isolated from all other gravitating bodies. As a consequence, the 4-dimensional general-relativistic spacetime around the Earth possesses a timelike Killing vector field. The (pseudo-)norm of this vector field is related to the first geoid which describes the gravitational redshift and, at the same time, the acceleration of falling corner cubes. The curl of this vector field gives a twist covector field. Outside of the Earth, where Einstein's vacuum field equation is assumed to hold, this covector field admits a potential which is known as the \emph{twist potential}, so we have a second potential related to the gravitational field of the Earth. The first potential is analogous to an electrostatic potential while  the second one is analogous to a magnetostatic potential. Accordingly, our two gravitational potentials can be regarded as a gravitoelectric and a gravitomagnetic potential, respectively. 

\section{The model of the Earth}

The Earth is an extended gravitating body. Therefore, it is most efficiently modeled in terms of a congruence of non-intersecting timelike worldlines \cite{Ehlers1993} describing the constituents of the Earth. As a first approximation, it is reasonable to assume that the Earth is rigidly rotating with a constant angular velocity. In this section we want to recollect the well-known fact that then the (appropriately parametrized) worldlines of the constituents are the integral curves of a timelike Killing vector field.  As 
most of the gravimetric measurements are taking place in the vacuum region outside of the Earth, what is important for us is the fact that the timelike Killing vector field that describes the motion of the constituents of the Earth can be extended, as a timelike Killing vector field, to the exterior region. The integral curves of this extended Killing vector field may be interpreted as the worldlines of geostationary satellites (or, {\color{blue} if} on the surface of the Earth, as the worldlines of observers that are at rest there with respect to the rotating Earth).

For a congruence of timelike curves with four-velocity $u^{\mu}$ one defines acceleration $a^{\mu}$, rotation $\omega _{\mu \nu}$, expansion $\theta$ and shear $\sigma _{\mu \nu}$ by the equations \cite{Ehlers1993}
\begin{equation}
\begin{split}
a^{\mu}= u^{\nu}D_{\nu} u^{\mu} \, , \quad
\omega _{\mu \nu} = D _{[\mu} u_{\nu ]} - a_{[\mu} u_{\nu ]} \, , \\
\theta = D_{\nu} u^{\nu} \, , \quad
\sigma _{\mu \nu} = D _{(\mu} u_{\nu )} - \dfrac{\theta}{3} h_{\mu \nu} - a_{(\mu} u_{\nu )} \end{split}
\end{equation}
where $h^\mu_\nu = \delta^\mu_\nu - u^\mu u_\nu$ is the projection onto the local rest space and $D_{\mu}$ is the covariant derivative defined by the Levi-Civita connection of the metric. Round brackets and square brackets denote symmetrization and antisymmetrization, respectively. In the following we discuss the operational meaning of these quantities for the special congruence associated with the rigidly rotating Earth. To that end we need the notion of a standard clock and of the radar distance between neighboring observers. 

Within General Relativity it is possible to uniquely characterize a particular parameter along the worldline of an observer which is called \textit{proper time}. An (idealized) clock that shows proper time is called a \textit{standard clock}. To give an operational characterization of this notion, we first define the \emph{radar distance} $\Delta x$ of an event $p$ from the worldline of a fixed observer. In our units with $c =1$, we simply have $\Delta x = \Delta t$ were $\Delta t$ is half the time span, measured with a clock of the observer, a light ray needs to propagate from the observer's worldline to $p$ and back to the observer. The observer's clock is a standard clock, i.e., the parameter $t$ is proper time, if and only if $\big( 1- (dx/dt)^2 \big) ^{-1} d^2x/dt^2 $ takes the same value for all freely falling particles emitted in the same spatial direction, see \cite{Perlick87}. It has been shown that the energy levels of atoms are influenced by the spacetime curvature according to $\delta E \sim R a_B^2$ where $R$ is a typical component of the curvature tensor and $a_B$ is the Bohr radius \cite{ParkerPimentel1982}. On Earth, this will amount to a relative frequency change of the order $\delta\nu/\nu \sim 10^{-42}$ which is more than 20 orders of magnitude beyond the present uncertainty of atomic clocks. Accordingly, with very high precision atomic clocks on Earth are standard clocks. With standard clocks it is also possible to uniquely define a standard distance. 

Furthermore, using light rays it is also possible to operationally define whether a congruence of timelike curves is rotating. This is Pirani's bouncing photon construction \cite{Pirani65}. Fix any two infinitesimally neighboring curves $A$ and $B$ in the congruence and send a light ray from $A$ to $B$. Reflect the light ray at $B$ in such a way that the tangent vectors to the incoming light ray, to the reflected light ray and to the worldline of $B$ are linearly dependent. We say that the congruence is irrotational if, in any such situation, the reflected light ray arrives back at $A$, i.e., if the light rays bouncing back and forth between $A$ and $B$ form a two-dimensional timelike worldsheet. It is well known that a congruence is irrotational if and only if it is hypersurface-orthogonal. In this case for any pair of infinitesimally neighboring worldlines $A$ and $B$ in the congruence the following is true: The normalized connecting vector $r^{\sigma}$ from $A$ to $B$, which is assumed to be orthogonal to the four-velocity vector $u^{\mu}$ tangent to the wordline of $A$, satisfies the Fermi(-Walker)-transport law $h^\mu_\nu u^\sigma D_\sigma r^\nu = 0$. Any deviation from that describes a rotation. This notion of rotation is often discussed  in the relativistic theory of continua, see e.g. \cite{Ehlers1993}, but it is well-defined also for congruences of worldlines in vacuum. 

As already mentioned, we want to assume, as a reasonable first approximation, that the Earth is rigid. In relativity a congruence is called (Born-)rigid if the (radar) distance between any two infinitesimally close worldlines of the congruence is time-independent. As a consequence, also angles between directions to neighboring worldlines remain constant in time. This is possible only if the congruence has vanishing shear and expansion \cite{Ehlers1993}. The rigidity condition still allows the Earth to rotate with $\omega_{\mu\nu} \neq 0$ and accelerate with $a^\mu \neq 0$. We assume now in addition that an observer co-moving with a constituent of the Earth always experiences the same situation. This means, in particular, that the acceleration of this co-moving observer is co-rotating, $h^\mu_\nu u^\rho D_\rho a^\nu = \omega^\mu{}_\nu a^\nu$. Furthermore, if we assume that the angular velocity of the rigidly rotating Earth is time-independent,  the rotation of the rigid Earth is assumed to be Fermi-constant, that is, $h_\rho^\kappa h^\sigma_\lambda u^\mu D_\mu \omega^\rho{}_\sigma = 0$. These three conditions which are fulfilled by the Earth to high precision then imply that the congruence describing the Earth is a Killing congruence, that is, $u \sim \xi$ with $\xi$ being a timelike Killing vector field \cite{Ehlers1993}. 

In reality, the Earth experiences small deformations, tides, winds, ocean whirls, snow falls and ice melting, and further time-dependent processes. All this happens with very low velocities and small masses so that the time-dependency of the gravitational field can be treated adiabatically to very high precision.

To sum up, with high precision the Earth is described adiabatically by means of a Killing congruence. This Killing congruence can then be extended to the exterior of the Earth. In this approximation, the analysis of the relativistic gravitational field of the Earth is thus tantamount to the analysis of a Killing congruence. In spherical polar coordinates $(t, r , \vartheta , \varphi )$ the Killing vector field is represented as $\xi = \partial _t + \Omega \partial _{\varphi}$ with a constant $\Omega$ that gives the angular velocity of the rotating Earth. Note that neither $\partial _t$ nor $\partial _{\varphi}$ are Killing vector fields, unless we assume that the Earth is axisymmetric. (The axisymmetric case, where we have an entire family of Killing vector fields parametrized by $\Omega$, will be treated in Section \ref{sec:axisymmetry} below.)  Strictly speaking, such an irregularly shaped rotating body would emit gravitational waves which would cause the angular velocity to decrease. However, for the Earth and all other planets and moons this energy loss by gravitational waves is totally negligible.

\section{The geometry of Killing congruences}

A Killing congruence is given by a timelike Killing vector field $\xi$ proportional to the 4-velocity of a family of observers, $\xi \sim u$. A Killing vector field possesses a (pseu\-do-)norm, $e^{2\phi} := g_{\mu\nu} \xi^\mu \xi^\nu$ as well as a curl $\partial_{[\mu} \xi_{\nu]}$ which is equivalent to the twist vector field $\varpi^\mu = \epsilon^{\mu\nu\rho\sigma} \xi_\nu \partial_\rho \xi_\sigma$. Here $\epsilon^{\mu\nu\rho\sigma} $ denotes the totally antisymmetric Levi-Civita tensor field (or volume form) associated with the spacetime metric where we choose the orientation such that in the spherical polar coordinates used below $\epsilon _{t r \vartheta \varphi} > 0$. The twist vector is Fermi(-Walker) propagated, $h^\mu_\nu u^\sigma D_\sigma \varpi^\nu = 0$. 

Using the Killing property of $\xi$, it can be shown that the (negative) gradient of the scalar function $\phi$ is equal to the acceleration of the Killing congruence, $a_\mu = - \partial_\mu \phi$ \cite{Ehlers1993}, which fulfills the co-moving condition $h^\mu_\nu u^\sigma D_\sigma a^\nu = \omega^\mu{}_\nu a^\nu$. If the metric satisfies Einstein's vacuum field equation, it can be shown that the twist covector field $\varpi_\mu = g_{\mu \nu} \varpi ^{\nu}$ possesses a potential, $\varpi_\mu = \partial_\mu \varpi$ \cite{IsraelWilson1972}. Therefore, in vacuum, outside the Earth, we have two gravitational potentials, $\phi$ and $\varpi$. Owing to their properties, see also below, these two potentials may be called \textit{gravitoelectric} and \textit{gravitomagnetic} potentials. It was already mentioned that they are the gravitational analogues of the electrostatic and magnetostatic potentials which are well-known in standard electromagnetism \cite{Jackson1999}.  

Up to an additive constant, each of the two gravitational potentials is directly related to measurements, either through a measurement of potential differences or through a measurement of the gradient. This will be outlined in the following. 

In our case of a stationary spacetime the line element can be 3+1 decomposed according to
\begin{equation}
ds^2 = e^{2\phi} \left(dt + \sigma_i dx^i\right)^2 - e^{-2\phi} \gamma_{ij} dx^i dx^j \, , \label{metricdecomp}
\end{equation}
with
\begin{equation}
\varpi^i = e^{4\phi} \varepsilon^{ijk} \partial_j \sigma_k
\end{equation}
where 
$i, j = 1, 2, 3$ and $\epsilon^{ijk}$ is the volume form associated with the spatial metric $\gamma_{ij}$ \cite{SimonBeig1982}. It is the $g_{00}$-component and the $g_{0i}$-component which can be expressed through the gravitoelectric and -magnetic potentials.

\section{The physics of the general-relativistic ``gravitoelectric'' potential of the Earth}

The gravitoelectric potential $\phi$ is obtained from the equation $e^{2 \phi} = g_{\mu\nu} \xi^\mu \xi^\nu$, where $\xi$ is assumed to be a timelike Killing vector field. As outlined above, we model the Earth as a rigid body that rotates with constant angular velocity; this allows us to choose spherical polar coordinates $(t, r, \vartheta , \varphi )$ in the vacuum region outside of the Earth such that $\xi = \partial _t + \Omega \partial _{\varphi}$, with a constant $\Omega$ that is to be identified with the angular velocity of the Earth.  The gravitoelectric potential foliates the spacetime into 3-dimensional hypersurfaces $\phi = \mathrm{const.}$.  Because of the Killing property of $\xi$, these hypersurfaces project onto 2-dimensional surfaces in the 3-dimensional space of integral curves of $\xi$ which are known as \emph{isochronometric surfaces}. The general-relativistic geoid can be defined as one of these surfaces, see Philipp et al. \cite{Philippetal2017}. In the case of the Earth it is natural to choose the isochronometric surface that is closest to the mean sea level. This surface is also used as a height reference, i.e., it has by definition zero height. Note that this definition of the geoid in terms of isochronometric surfaces makes sense not only for the Earth but also for all other planets and moons and even for neutron stars and black holes. In the latter cases the choice of a particular isochronometric surface is purely conventional. The application to compact and ultracompact objects is made possible by the fact that exact relativistic equations are used, rather than post-Newtonian approximations. The idea of defining the geoid in terms of isochronometric surfaces was brought forward already in 1985 by Bjerhammar \cite{Bjerhammar1985} who, however, did not work out any mathematical details. Also inspired by Bjerhammar, Kopeikin et al. \cite{KopeikinMazurovaKarpik2015} discussed a relativistic geoid based on a particular fluid model of the Earth. For an alternative fully relativistic definition of a geoid, not in general related to an operational realization with clocks, we  refer to Oltean et al. \cite{OlteanEtAl2015}.  

The gravitoelectric potential can be most easily measured and, thus, operationally defined through the redshift in clock-comparison experiments \cite{Philippetal2017}: For any two stationary clocks (i.e., clocks whose  worldlines are integral curves of the Killing vector field $\xi$) the redshift $z$ is given by the equation 
$\mathrm{ln} (1+z) = \phi _2 - \phi _1 $ where $\phi _1$ and $\phi _2$ are the values of $\phi$ at the positions of the two clocks. (Note that $\phi$ is constant along each worldline of $\xi$.) As outlined in \cite{Philippetal2017}, for clocks on the surface of the Earth the comparison may be done with the help of optical fibers. Such fiber links already exist and they may have a length of more than thousand kilometers \cite{Koke:2019bwo}. As an alternative to redshift measurements, the same potential difference $\phi _2 - \phi _1$ can also be obtained with atom interferometry \cite{dosSantos2016,Audretsch:1982ux,Kagramanova:2008bv}. From measurements of the acceleration $a_\mu = - \partial_\mu \phi$ of a falling corner cube  one can also calculate the equipotential surfaces. Therefore, within full General Relativity, all three types of measurements yield the \textit{same} potential $\phi(x)$ which makes data fusion and improved geoid determination possible \cite{Wu2019}.  

\section{The physics of the general-relativistic ``gravitomagnetic'' potential of the Earth}

There are many ways to determine the gravitomagnetic potential through measurements of the twist potential. 

(i) The Sagnac effect for light is sensitive to the twist. The Sagnac experiment runs with a ring laser interferometer with counter-propagating laser beams. The two interfering beams give a proper time difference given by \cite{AshtekarMagnon75}
\begin{equation}
\Delta t = 2 e^{\phi_0} \oint e^{- \phi} \xi_i dx^i
\end{equation}
where $dx^i$ is the spatial differential within the interferometer's rest frame. Using Stokes's theorem and the twist vector this can be rewritten as
\begin{equation}
\Delta t = e^{\varphi_0} \int_\Sigma e^{-3\phi} \epsilon_{ijk} \varpi^k d\Sigma^{ij} \approx e^{-2\phi_0} \vec\varpi \cdot \vec\Sigma\, ,
\end{equation}
for a small interferometer with area $\vec\Sigma$ and $\phi_0$ as gravitoelectric potential at the position of the beam splitter. If the Einstein vacuum field equations are fulfilled, then we introduce $\Phi_J = e^{-2\phi} \varpi$ which is the angular momentum potential and obtain
\begin{equation}
\Delta t = \nabla\Phi_J \cdot \vec\Sigma\, .
\end{equation}
Similar characterizations of the gravitomagnetic potential result from atom interferometry \cite{Audretsch:1982ux}.

(ii) Also the propagation of classical objects with spin $S^\mu$ couple to the gravitomagnetic field. This is known as the Schiff effect first derived in \cite{Schiff1960}. The relation to the twist has been shown in \cite{Zimbres:2014toa}
\begin{equation}
h^\mu_\nu u^\sigma D_\sigma S^\mu = e^{- 2 \phi} \epsilon^{\mu\nu}{}_{\rho\sigma} u^\rho \varpi_\nu S^\sigma  \, , 
\end{equation}
see also \cite{RindlerPerlick1990}. This effect has experimentally been confirmed by the space mission Gravity Probe B \cite{Everittetal12}. 

(iii) Finally, atomic spectroscopy is also sensitive to the Sagnac effect \cite{Silverman1991}. 

All these methods measure the same twist and, thus, the same gravitomagnetic potential. 

\section{Stationary and axisymmetric spacetimes}\label{sec:axisymmetry}

In order to get a better physical and intuitive understanding of these potentials we are now calculating and visualizing the gravitoelectric and the gravitomagnetic potential for certain examples of spacetimes. For doing so, we now specialize to the case that the spacetime is stationary and axisymmetric. In this case we can use spherical polar coordinates $(t,r,\vartheta , \varphi )$ such that $\partial _t$ and $\partial _{\varphi}$ are Killing vector fields. We can then consider the gravitoelectric and gravitomagnetic potentials with respect to the Killing vector field $\partial _t + \Omega \partial _{\varphi}$, with \emph{any} constant $\Omega$. The potentials are well-defined on the domain where $\partial _t + \Omega \partial _{\varphi}$ is timelike. Because of the symmetry they are functions of $r$ and $\vartheta$ only.

For modeling the gravitational field around the Earth the assumption of axisymmetry
is of course a strong idealization. The realistic Earth can be modeled using a relativistic multipole expansion.

If the stationary and axisymmetric spacetime is asymptotically flat (i.e., 
if $g_{tt} \to 1$ and $g_{\varphi \varphi}/(r^2 \mathrm{sin} ^2 \vartheta )  
\to -1$ for $r \to \infty$) , the Killing vector field $\partial_t$ is 
distinguished among the Killing vector fields $\partial_t +
\Omega \partial_{\varphi}$ by the property that it is timelike 
near infinity. The integral curves of $\partial _t$ can then be 
interpreted as the worldlines of observers who co-rotate with the 
source. The gravitoelectric and gravitomagnetic potentials related
to this Killing vector field are commonly combined to give the 
complex \emph{Ernst potential}
\begin{equation}\label{eq:Ernst}
\mathcal{E} = e^{2 \phi} + i \varpi \, .
\end{equation}
Einstein's vacuum field equation then reduces to a partial differential equation for 
$\mathcal{E}$ known as the \emph{Ernst equation}, see e.g. Griffiths and Podolsk{\' y}
\cite{GriffithsPodolsky2009} for details. The Ernst potential determines all components of the metric. In other words, a stationary axisymmetric vacuum spacetime is completely determined by the two real potentials $\phi$ and $\varpi$ which depend only on the two coordinates $r$ and $\vartheta$. This should be contrasted with an arbitrary spacetime, where the metric has ten independent components which depend on all four coordinates, and with a stationary vacuum spacetime, where the metric is determined by eight scalar-valued functions that depend on the three spatial coordinates.

The idea of using the real 
and the imaginary part of the Ernst potential as coordinates was brought forward 
already in the 1980s by Perj{\'e}s \cite{Perjes1988}. These two real potentials
coordinatize the planes $(t, \varphi ) = \mathrm{const.}$, i.e., together with
$t$ and $\varphi$ they can be used as coordinates on the spacetime. Of course, 
this is true only where $d \phi$ and $d \varpi$ are linearly independent. Whereas the
potentials corresponding to $\partial _t + \Omega \partial _{\varphi}$ with $\Omega = 0$
are the ones most naturally related to the spacetime geometry, these potentials can 
be defined with respect to any $\Omega$. This is of particular relevance in the case 
that the source is non-rotating, i.e., that $\partial _t$ is hypersurface-orthogonal. 
Then the gravitomagnetic potential associated with $\Omega = 0$ is constant and 
cannot be used as a coordinate.

If the metric is given as
\begin{equation}
ds^2 =
g_{tt} dt^2 + g_{rr} dr^2 + g_{\vartheta\vartheta} d\vartheta^2 + g_{\varphi\varphi} d\varphi^2 + 2 g_{t\varphi} dt \, d\varphi \, ,
\end{equation}
with the $g_{\mu \nu}$ depending only on $r$ and $\vartheta$, the gravitoelectric 
potential associated with the Killing vector field $\partial _t + 
\Omega \partial _{\varphi}$ is given by the equation
\begin{equation}
e^{2 \phi} = g_{tt} + 2 \Omega g_{t\varphi} + \Omega^2 g_{\varphi\varphi} 
\, .
\label{eq:phiaxi}
\end{equation}
The twist covector field reads
\[
\varpi_{\mu} dx^{\mu} = 
\frac{g_{rr} g_{tt}^{\: 2}}{\sqrt{-g}} 
\Bigg( \partial_\vartheta \dfrac{g_{t\varphi}}{g_{tt}} + \Omega \partial_\vartheta \dfrac{g_{\varphi\varphi}}{g_{tt}} - \Omega^2 \dfrac{g_{\varphi \varphi}^{\: 2}}{g_{tt}^{\: 2}} \partial_\vartheta \dfrac{g_{t \varphi}}{g_{\varphi\varphi}}\Bigg) dr 
\]
\begin{equation}
- \frac{g_{\vartheta \vartheta} g_{tt}^{\: 2}}{\sqrt{-g}} 
\Bigg( \partial_r \dfrac{g_{t\varphi}}{g_{tt}} + \Omega \partial_r \dfrac{g_{\varphi\varphi}}{g_{tt}} - \Omega^2 \dfrac{g_{\varphi \varphi}^{\: 2}}{g_{tt}^{\: 2}} \partial_r \dfrac{g_{t \varphi}}{g_{\varphi\varphi}}\Bigg) d \vartheta 
\label{eq:piaxi}
\end{equation}
with $g = \left(g_{tt} g_{\varphi\varphi} - g_{t\varphi}^2\right) g_{rr} g_{\vartheta\vartheta}$. If the metric satisfies Einstein's vacuum field equation, it is guaranteed that the twist covector field admits a potential, $\varpi _{\mu} = \partial _{\mu}\varpi$. We now discuss the gravitoelectric and gravitomagnetic potentials for a few specific stationary and axi\-symmetric vacuum solutions to Einstein's field equation, 
thereby illustrating that they can be applied also to the spacetime around a black 
hole or another (ultra-)compact object. In all cases, we consider the metric in spherical 
polar coordinates $(t, r, \varphi , \vartheta)$, where $\partial _t$ and $\partial _{\varphi}$ 
are Killing vector fields. We determine the gravitoelectric and gravitomagnetic 
potentials with respect to the Killing vector field $\partial _t + \Omega \partial _{\varphi}$
and we plot the potentials $\phi (r, \vartheta )$ and $\varpi (r , \vartheta )$ in diagrams
with $r \, \mathrm{sin} \, \vartheta$ on the horizontal axis and $r \, \mathrm{cos} \, \vartheta$
on the vertical axis. The examples demonstrate how the two families of equipotential lines give a coordinatization of the $r$-$\vartheta$-plane. For sufficiently small $|\Omega |$ the gravitoelectric equipotential lines close to the central body are circles, i.e., the gravitoelectric geoid in 3-space has the topology of a sphere, as usually associated with the term ``geoid''. By contrast, the gravitomagnetic equipotential lines are not usually closed, i.e., the gravitomagnetic geoid in 3-space has the topology of $\mathbb{R}^2$. To put this another way, the gravitoelectric potential may be interpreted as a height coordinate whereas the gravitomagnetic potential may be viewed as a latitude coordinate.

\subsection{Schwarzschild, Kottler and Reissner-Nordstr{\"o}m metrics}

For a spherically symmetric and static metric of the form
\begin{equation}
    ds^2 = f(r)  dt^2 
    - \dfrac{dr^2}{f(r)}
    - r^2 \Big( d \vartheta ^2 + \mathrm{sin} ^2 \vartheta \, d \varphi ^2 \Big)
\end{equation}
(\ref{eq:phiaxi}) gives us the gravitoelectric potential as
\begin{equation}
e^{2 \phi} = f(r) - \Omega^2 r^2 \mathrm{sin} ^2\vartheta 
\end{equation}
and (\ref{eq:piaxi}) gives us the twist covector field as
\begin{equation}
\varpi_\mu dx^{\mu}=  \Omega  \Big( 2 \cos\vartheta \, dr
 - r \sin\vartheta  \big( 2 f(r)-r f'(r) \big) \, d \vartheta \Big) \, .  
\end{equation}
For the Scharzschild spacetime,
\begin{equation}
    f(r) = 1 - \dfrac{2M}{r} \, ,
\end{equation}
this specifies to 
\begin{equation}
e^{2 \phi} = 1 - \dfrac{2M}{r} - \Omega^2 r^2 \mathrm{sin} ^2\vartheta 
\end{equation}
and 
\begin{equation}
\varpi_\mu dx^{\mu}=  2 \Omega  \Big( \cos\vartheta \, dr
 -  \big( r- 3M \big) \, \sin\vartheta  \, d \vartheta  \Big)\, . 
 \label{eq:pimuSchw}
\end{equation}
Integration of the latter equation gives us the gravitomagnetic potential
\begin{equation}
\varpi = 2 \Omega (r - 3 M) \cos\vartheta \, .
\label{eq:piSchw}
\end{equation}
More generally, we can consider the Kottler spacetime, also known as the Schwarzschild-(anti)deSitter spacetime, 
\begin{equation}
    f(r) = 1 - \dfrac{2M}{r} - \dfrac{\Lambda r^2}{3}  \, .
\end{equation}
In this case the cosmological constant gives a contribution to the gravitoelectric potential,   
\begin{equation}
e^{2 \phi} = 1 - \dfrac{2M}{r} - \dfrac{\Lambda r^2 }{3}- \Omega^2 r^2 \mathrm{sin} ^2\vartheta \, , 
\end{equation}
whereas it drops out from the equation for the twist co\-vec\-tor field, which is still given by (\ref{eq:pimuSchw}). So also in this case we have a gravitomagnetic potential given 
by (\ref{eq:piSchw}).

However, in the case of the Reissner-Nordstr\"om spacetime
\begin{equation}
    f(r) = 1 - \dfrac{2M}{r}+ \dfrac{Q^2}{r^2}
\end{equation}
the twist covector field 
\begin{equation}
\varpi_\mu d x ^{\mu} 
= 
2 \Omega 
\Bigg(  
\cos\vartheta \, dr 
-
\Big(r - 3 M + 2 \frac{Q^2}{r}\Big) \sin\vartheta \, d \vartheta 
\Bigg) 
\end{equation}
is not integrable (unless $\Omega = 0$ or $Q = 0$ ), i.e., in this case the gravitomagnetic potential does not exist. This is in line with the theorem in \cite{IsraelWilson1972}. 

\begin{figure*}[t]
\subfigure[][Schwarzschild with $\Omega = 0.18 M^{-1}$]{\includegraphics[width=0.26\textwidth]{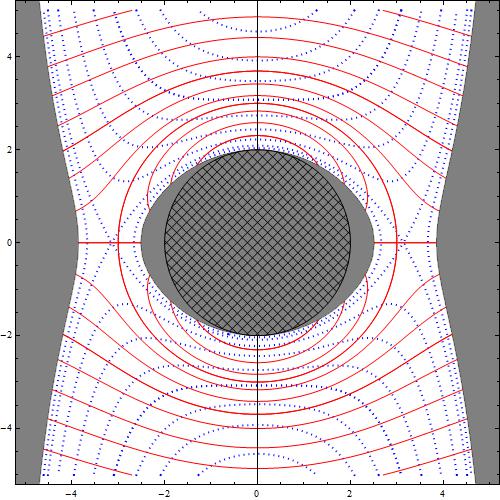}}
\subfigure[][Schwarzschild with $\Omega = 0.20 M^{-1}$]{\includegraphics[width=0.26\textwidth]{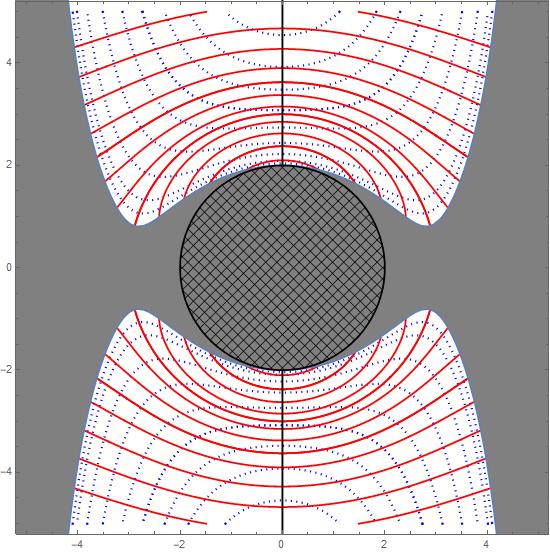}}
\caption{The gravitoelectric (blue dotted lines) and gravitomagnetic (red lines) potentials for a Schwarzschild spacetimes. The crosshatched area is the region inside the black-hole horizon and the gray shaded area is the region where the potentials are not defined because the Killing vector field is spacelike. \label{fig:ex1a}}
\end{figure*}

The equipotential surfaces for Schwarzschild are shown in Fig.~\ref{fig:ex1a} (a) and (b) for two different values of the angular velocity $\Omega$ of the observer field. (For $\Omega =0$ we have of course $\varpi = 0$, so in this case the potentials cannot be used as coordinates.)
There is a forbidden region, shown in gray shading, where the potentials are not defined because the Killing vector field $\partial _t + \Omega \partial _{\varphi}$ is spacelike. Outside of this region, the differentials $d\phi$ and $d \varpi$ are linearly independent, for all non-zero values of $\Omega$, with the exception of the axis. If $|\Omega |$ is smaller than the critical value $\Omega _{\mathrm{crit}} = (\sqrt{27} M)^{-1} \approx 0.19245 M^{-1}$, the forbidden region consists of two connected components: One is adjacent to the horizon, the other one is the region outside of the so-called light-cylinder.  The inner equipotential surfaces of the gravitoelectric potential are topological spheres in 3-space, the outer ones are topological cylinders; the borderline case is an equipotential surface with a self-crossing along a circle in the equatorial plane. The equipotential surfaces of the gravitomagnetic potential all have the topology of $\mathbb{R}^2$, with the exception of one which consists of the sphere $r=3M$ and part of the equatorial plane. If $| \Omega |= \Omega _{\mathrm{crit}}$ the two connected components of the forbidden region touch at the photon circle $r=3M$ in the equatorial plane. If $| \Omega | > \Omega _{\mathrm{crit}}$, the forbidden region is connected. Now all the equipotential surfaces have the topology of $\mathbb{R}^2$ in 3-space. 

For the Earth the critical angular velocity is $\Omega_{\text{Earth},\text{crit}} \sim 800 \;\text{GHz}$ which is 16 orders of magnitude larger than the actual value. The outer part of the forbidden region begins at approx. $5 \times 10^9\;\text{km}$ which is 10 times the distance to Pluto. And the inner part of the forbidden region does not exist for the real Earth. 

\begin{widetext}
\subsection{Kerr metric}

We now consider the Kerr metric
\[
    ds^2 =
 \left( 1-\dfrac{2Mr}{r^2 + a^2 \mathrm{cos} ^2 \vartheta} \right) dt^2 
 - \dfrac{r^2+a^2 \mathrm{cos} ^2 \vartheta}{r^2-2 M r +a^2} dr^2 
-(r^2+ a^2 \mathrm{cos} ^2 \vartheta ) d\vartheta^2 
\]
\begin{equation}
- \sin^2 \vartheta \Big(r^2 + a^2 + \dfrac{2Mra^2\sin^2\vartheta}{r^2+ a^2 \mathrm{cos} ^2 \vartheta} \Big) d\varphi^2
+ \dfrac{4Mra\mathrm{sin} ^2\vartheta}{r^2+ a^2 \mathrm{cos} ^2 \vartheta} dt \, d\varphi
\end{equation}
where, for a Kerr black hole, $a^2 \le M^2$.
By (\ref{eq:phiaxi}) the gravitoelectric potential reads
\begin{equation}
e^{2 \phi} = 
1 - \frac{2 M r}{r^2 + a^2 \cos^2\vartheta} + 
\Omega \frac{4 M a r \sin^2\vartheta}{r^2 + a^2 \cos^2\vartheta} 
- 
\Omega^2 \left(r^2 + a^2 + \frac{2 M a^2 r \sin^2\vartheta}{r^2 + a^2 \cos^2\vartheta}\right) \sin^2\vartheta  \, . 
\end{equation}
From (\ref{eq:piaxi}) we find the twist covector field 
\begin{equation}
\varpi_\mu dx^{\mu} 
= 
\frac{\left(\varpi^{(0)}_\mu + \Omega \varpi^{(1)}_\mu + \Omega^2\varpi^{(2)}_\mu\right) \, dx^{\mu}}{\left(r^2 + a^2 \cos^2\vartheta\right)^2} 
\end{equation}
with
\begin{eqnarray}
\varpi^{(0)}_\mu dx^{\mu} & = & 
-2 a M \Bigg( 2 r \cos\vartheta \, d r 
+
\left(r^2 - a^2 \cos^2\vartheta\right) \sin\vartheta \, d \vartheta \Bigg) \\
\varpi^{(1)}_\mu d x ^{\mu} 
& = & 
2 \cos\vartheta \Big((r^2 + a^2)^2 
- 2 a^2 (r^2 + a^2 - 2 M r) \mathrm{sin} ^2\vartheta + a^4 \mathrm{sin} ^4 \vartheta\Big) dr \nonumber\\ 
& & -
2 \sin\vartheta \Big((r^2 + a^2) \left(r^2 (r - 3 M) + a^2 (M + r) - 2 a^2 r \, \mathrm{sin} ^2\vartheta\right) 
- a^4 (M - r) \mathrm{sin} ^4\vartheta\Big) d \vartheta \, , \\ 
\varpi^{(2)}_\mu dx ^{\mu}
& = & 
-2 \, a M \mathrm{sin} ^3 \vartheta
\Bigg(
2 a^2 r \cos\vartheta \sin \vartheta \, dr
- \Big(a^4 \cos^2\vartheta 
- a^2 r^2 \mathrm{sin} ^2 \vartheta 
- 3 r^4\Big) \, 
d \vartheta \Bigg)
\, . 
\end{eqnarray}
This gives us the gravitomagnetic potential
\begin{equation}
\varpi = \Big(1 + \Omega^2 \left(2 \left(r^2 + a^2\right) + \left(r^2 - a^2\right) \sin^2\vartheta\right)\Big) \frac{2 a M \cos\vartheta}{r^2 + a^2 \cos^2\vartheta} 
+ 2 \, \Omega \cos\vartheta \frac{\left(r^2 + a^2\right) (r - 3 M) - a^2 (r-M) \sin^2\vartheta}{r^2 + a^2 \cos^2\vartheta} 
\, . 
\end{equation}

\begin{figure*}[t]
\subfigure[][Kerr with $a=0.90 M$, $\Omega = 0$]{\includegraphics[width=0.26\textwidth]{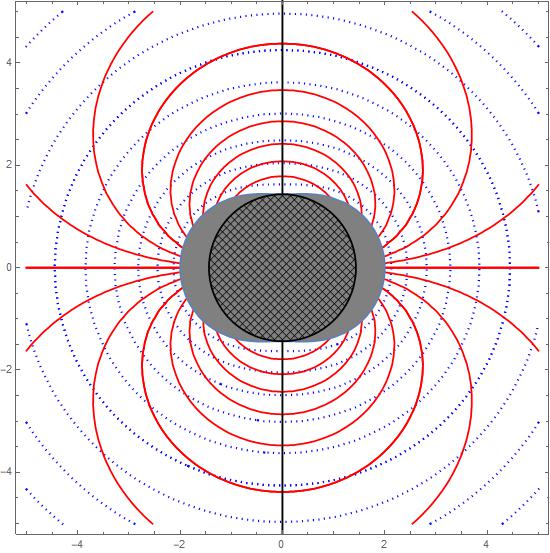}}
\subfigure[][Kerr with $a=0.90 M$, $\Omega = 0.20 M^{-1}$]{\includegraphics[width=0.26\textwidth]{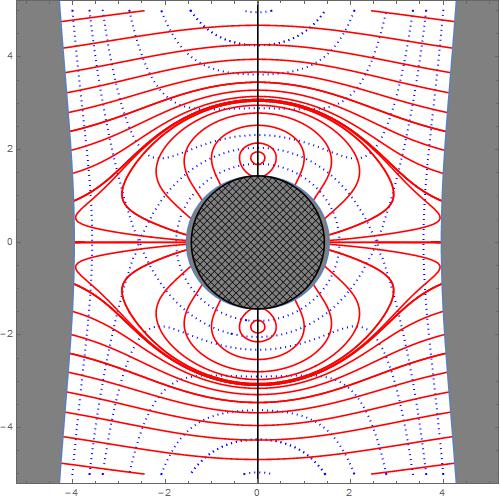}}
\subfigure[][Kerr with $a=0.90 M$, $\Omega = 0.362 M^{-1}$]{\includegraphics[width=0.26\textwidth]{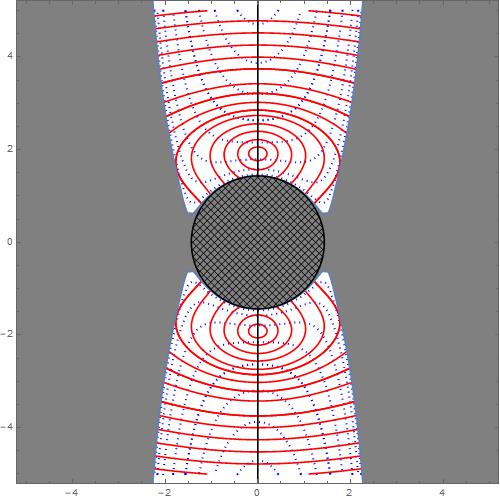}}
\subfigure[][Kerr with $a=0.90 M$, $\Omega = -0.143 M^{-1}$]{\includegraphics[width=0.26\textwidth]{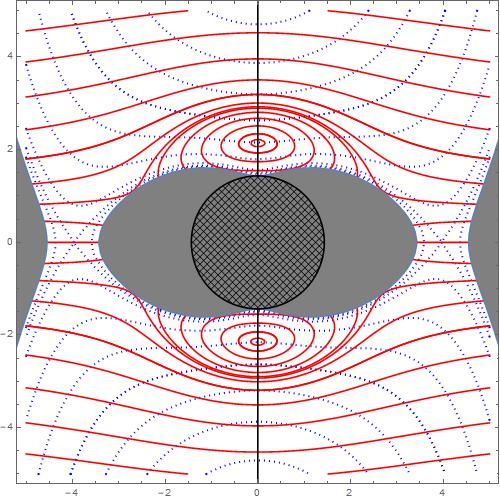}}
\subfigure[][Kerr with $a=0.90 M$, $\Omega = -0.147 M^{-1}$]{\includegraphics[width=0.26\textwidth]{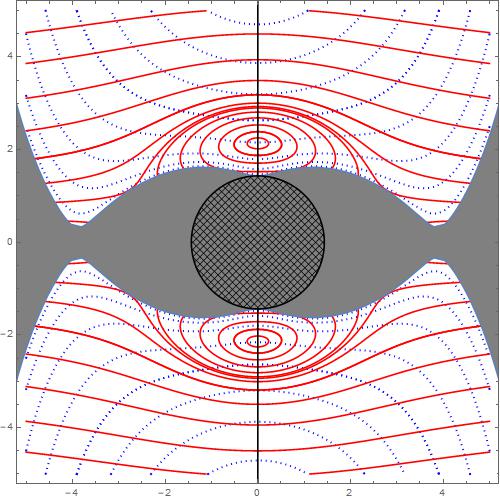}}
\caption{The gravitoelectric (blue dotted lines) and gravitomagnetic (red lines) potentials  for Kerr spacetimes. The crosshatched area is the region inside the outer horizon and the gray shaded area is the region where the potentials are not defined because the Killing vector field is spacelike.\label{fig:ex1b}}
\end{figure*}

In Fig.$\,$\ref{fig:ex1b}$\,$(a)  we show the equipotential surfaces for non-rotating observers, $\Omega =0$, where the potentials simplify to \cite{Hansen1974}
\begin{equation}
e^{2 \phi} = 
1 - \frac{2 M r}{r^2 + a^2 \cos^2\vartheta}  
\, , 
\end{equation}
\begin{equation}
\varpi = \frac{2 \, a M \cos\vartheta}{r^2 + a^2 \cos^2\vartheta} \, .
\end{equation} 
The surfaces $\phi = \mathrm{const.}$ are topological spheres in 3-space; the innermost one is the boundary of the ergoregion. The surfaces $\varpi = \mathrm{const.}$ all have the topology of $\mathbb{R}^2$. For rotating observers we have to distinguish the case $a \Omega >0$ (Fig.$\,$\ref{fig:ex1b}$\,$(b) and (c)) and the case $a \Omega < 0$ (Fig.~\ref{fig:ex1b}~(d) and (g)). If $|\Omega|$ is sufficiently small, the region where the potentials are not defined (gray shaded in the figure) is connected; beyond a critical value this region decomposes into two connected components which are separated from the equatorial plane. 

\subsection{The NUT metric}
The NUT metric \cite{NewmanUntiTamburino1963} is a solution to Einstein's vacuum field equation that reads
\begin{equation}
ds^2 = \dfrac{r^2-2 M r-n^2}{n^2+r^2} \Big( dt- 2 \, n \, \mathrm{cos} \, \vartheta \, d \varphi  \Big) ^2-\dfrac{(r^2+n^2) dr^2}{r^2-2 M r-n^2} -\big( r^2+n^2 \big)  \Big(d \vartheta ^2 + \mathrm{sin}^2 \vartheta  \, d \varphi ^2 \Big)
\end{equation}
where $M$ is the mass parameter and $n$ is the socalled NUT parameter, also known as a \emph{gravitomagnetic charge}. From (\ref{eq:phiaxi})
we find the gravitoelectric potential
\begin{equation}
e^{2 \phi}  =  
\dfrac{r^2-2Mr-n^2}{r^2+n^2} \, 
\big( 1 - 2 n \Omega \, \mathrm{cos}\, \vartheta \big) ^2
- \Omega ^2 (r^2+n^2) \, \mathrm{sin} ^2 \vartheta 
\end{equation}
and from (\ref{eq:piaxi}) we find the twist covector field
\begin{eqnarray}
\varpi _{\mu} dx^{\mu} & = & 
- \Bigg(
\Big( 1 - 2 \, n \, \Omega \, \mathrm{cos} \, \vartheta \Big) ^2
\dfrac{2 \, n \big( r^2-2Mr-n^2 \big)}{\big( r^2+n^2 \big) ^2} -
2 \Omega 
\Big( \mathrm{cos} \, \vartheta \big( 1 - n \, \Omega \, \mathrm{cos} \, \vartheta \big)
- n \, \Omega \Big) \Bigg) \, dr \nonumber\\
& & -2 \, \Omega \, \mathrm{sin} \, \vartheta \, 
\big( 1 - 2 \, n \, \Omega \, \mathrm{cos} \, \vartheta \big)  
\dfrac{r^3-3Mr^2-3n^2r+Mn^2}{r^2+n^2} \, d \vartheta \, .
\end{eqnarray}
The twist potential reads
\begin{equation}
    \varpi  =  \dfrac{2n(r-M)}{r^2+n^2} + 
2 \, \Omega \, \mathrm{cos} \, \vartheta 
\Big( 1 - n \, \Omega \, \mathrm{cos} \, \vartheta \Big)
\dfrac{r^3-3Mr^2-3n^2r+Mn^2}{r^2+n^2} -
n (2r+3M) \Omega ^2  
\end{equation}
Fig.$\,$\ref{fig:ex2}$\,$(a) and (b) show the potentials for the NUT metric with two different values of $\Omega$. Qualitatively, the equipotential surfaces are similar to the Schwarzschild case. In particular there is a critical value for $\Omega$ beyond which the region where the potentials are not defined decomposes into two connected components. Note, however, that in contrast to the Schwarzschild spacetime the potentials are no more symmetric with respect to the equatorial plane.

\subsection{The Kerr-NUT metric}

The Kerr-NUT metric is a solution to Einstein's vacuum field equation that depends on a mass parameter $M$, a NUT parameter (gravitomagnetic charge) $n$ and a spin parameter $a$, see e.g. Griffiths and Podolsk{\' y} \cite{GriffithsPodolsky2009}, p.312. The metric reads
\[
ds^2 = 
\dfrac{
(r^2+a^2-n^2-2Mr)
}{
r^2+(n-a \, \mathrm{cos} \, \vartheta )^2
}
\Big( dt - \big( 
a \, \mathrm{sin} ^2 \vartheta + 2 \, n \, \mathrm{cos} \, \vartheta 
\big) d \varphi \Big) ^2
-
\dfrac{
r^2+(n-a \, \mathrm{cos} \, \vartheta )^2
}{
(r^2+a^2-n^2-2Mr)
}
\, dr^2 
\]
\begin{equation}
-
\dfrac{\mathrm{sin} ^2 \vartheta }{r^2+(n-a \, \mathrm{cos} \, \vartheta )^2}
\Big( 
a \, dt - (r^2+a^2+n^2) d \varphi 
\Big) ^2
-
\Big( r^2 + (n-a \, \mathrm{cos} \, \vartheta )^2 \Big) d \vartheta ^2
\, .
\end{equation}
For simplicity, we restrict here to $\Omega = 0$. Then the gravitoelectric potential is
\begin{equation}
e^{2 \phi} = 1 - \frac{2 (Mr + n^2-a \, n \, \mathrm{cos} \, \vartheta)}{r^2+(n-a \, \mathrm{cos} \, \vartheta )^2} \, , 
\end{equation} 
and the twist covector field is
\[
\varpi_\mu = 
\frac{
4 M r \left(n - a \cos\vartheta\right) - 2 n \left(r^2 - \left(n - a \cos\vartheta\right)^2\right) 
}{
\left(r^2 + \left(n - a \cos\vartheta\right)^2\right)^2
} \, dr
\]
\begin{equation}
-
\frac{
2 \, a \Big(M \left(r^2 - (n - a \cos\vartheta)^2\right) + 2 r n (n - a \cos\vartheta)\Big) \sin\vartheta
}{
\left(r^2 + \left(n - a \cos\vartheta\right)^2\right)^2
} \, d \vartheta \, ,
\end{equation}
\end{widetext}
so the gravitomagnetic potential equals
\begin{equation}
\varpi = 2 \, \frac{ n r-M \left(n - a \cos\vartheta\right) }{r^2 + \left(n - a \cos\vartheta\right)^2} \, .
\end{equation}

The potentials for the Kerr-NUT metric with $\Omega = 0$ are shown in Fig.$\,$\ref{fig:ex2}$\,$(c) for the case $a n >0$ and in Fig.$\,$\ref{fig:ex2}$\,$(d) for the case $a n <0$. In either case, the surfaces $\phi = \mathrm{const.}$ are topological spheres in 3-space and the surfaces $\varpi = \mathrm{const.}$ are diffeomorphic to $\mathbb{R}^2$. Again, it is clearly seen that the NUT parameter breaks the symmetry with respect to the equatorial plane.

\subsection{The rotating $q$-metric}

The rotating $q$-metric is a solution to Einstein's vacuum field equation that was found by Toktarbay and Quevedo  \cite{Toktarbay:2015lua}, also see \cite{Frutos-Alfaro:2016arb}. It depends on 3 parameters, a mass parameter $M$, a quadrupole parameter $q$ and a spin parameter $a$. It features a naked singularity which is considered as unphysical by most authors. Therefore, when working with this metric one usually assumes that this vacuum solution is valid only outside of a certain sphere which covers the naked singularity and that inside this sphere the metric has to be matched to a regular interior solution. If interpreted in this sense the rotating $q$-metric describes the spacetime around a spinning body with a non-zero quadrupole moment that is very compact but not compact enough to have undergone gravitational collapse. As the surface where the matching is done can be chosen arbitrarily close to the naked singularity, in the following we consider the metric down to the naked singularity. 

As the $q$-metric has a non-zero quadrupole moment, it is of geodetic relevance since it might be used to relativistically model the flattened Earth. Higher-order multipole moments can be introduced via a series expansion of the metric, see e.g., \cite{SimonBeig1982}.

The Ernst potential $\mathcal{E}$ for the rotating $q$ metric can be found in  \cite{Toktarbay:2015lua}. From that we can find for non-rotating observers, $\Omega = 0$,  the gravitoelectric and gravitomagnetic potentials via (\ref{eq:Ernst}). In prolate spheroidal coordinates $(t,x,y,\varphi )$, the Ernst potential reads
\begin{equation}
\mathcal{E} = \left( \dfrac{x-1}{x+1} \right) \left( \dfrac{x-1+(x^2-1)^{-q} d_+}{x+1+(x^2-1)^{-q} d_-} \right)
\end{equation}
where
\begin{eqnarray}
d_{\pm} & = & -\alpha ^2 (x \pm 1) h_+ h_- (x^2-1)^{-q} \nonumber\\ 
& & + i \alpha \Big(y (h_+ + h_- ) \pm (h_+ - h_-)\Big) \\
h_{\pm} & = & (x \pm y)^{2q} \\ 
\alpha a & = & \sigma - m \, .
\end{eqnarray}
The prolate spheroidal coordinates are related to spherical polar coordinates $(t, r , \vartheta , \varphi )$ by the transformation
\begin{equation}
x = \dfrac{r- M}{\sigma} \, , \quad 
y = \mathrm{cos} \, \vartheta \, .
\end{equation} 
Here $\sigma$ is a constant parameter. For rotating observers ($\Omega \neq 0$) the potentials are given by very involved equations which will not be written out here but can be easily evaluated numerically.   

\begin{figure*}[t]
\subfigure[][NUT with $n=0.50 M$, $\Omega = 0.175 M^{-1}$]{\includegraphics[width=0.26\textwidth]{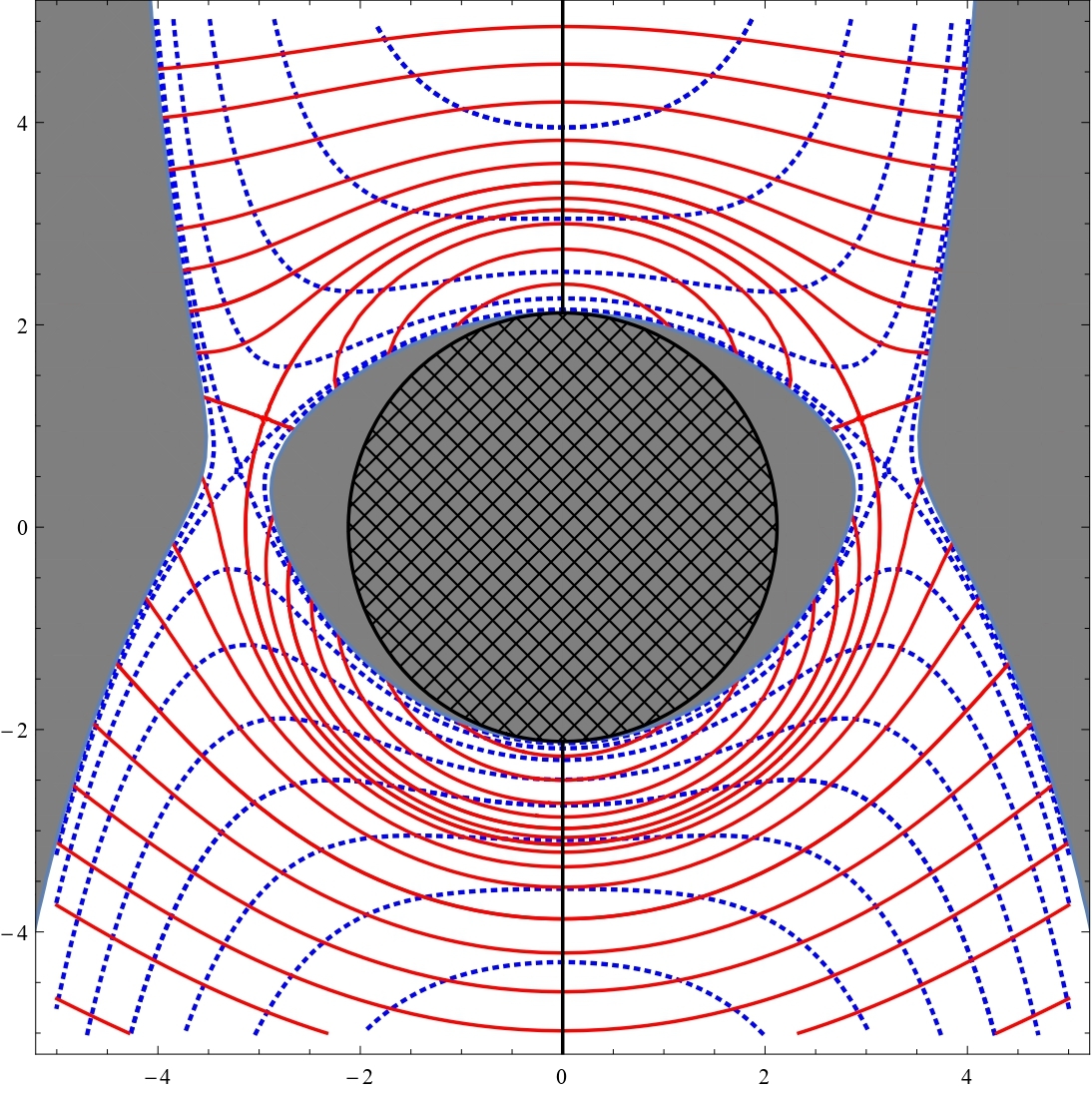}}
\subfigure[][NUT with $n=0.50 M$, $\Omega = 0.185 M^{-1}$]{\includegraphics[width=0.26\textwidth]{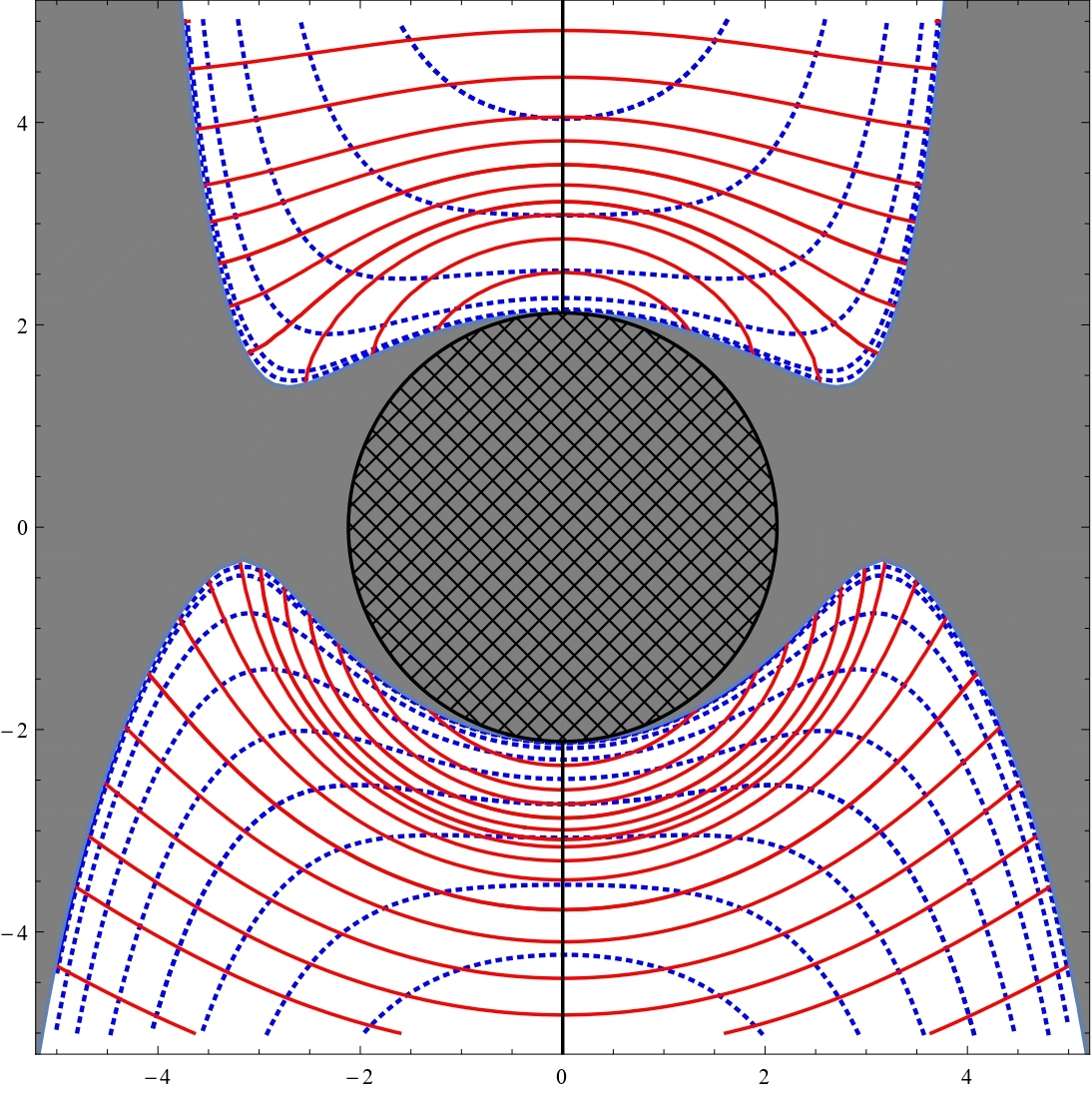}}
\subfigure[][Kerr-NUT with $a=0.90 M$, $n=0.50 M$, $\Omega = 0$]{\includegraphics[width=0.26\textwidth]{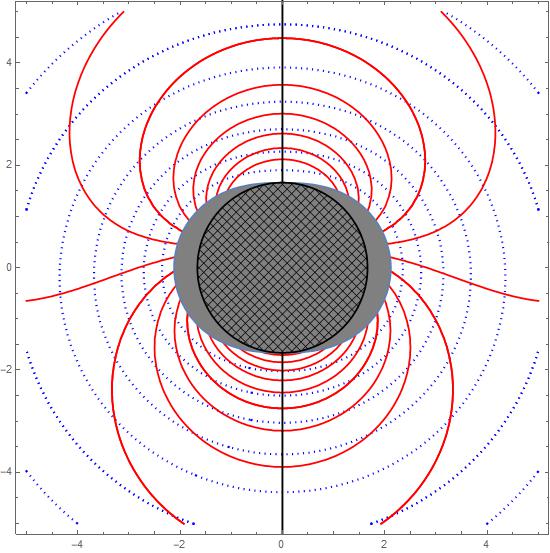}}
\subfigure[][Kerr-NUT with $a=0.90 M$, $n=-0.50 M$, $\Omega = 0$]{\includegraphics[width=0.26\textwidth]{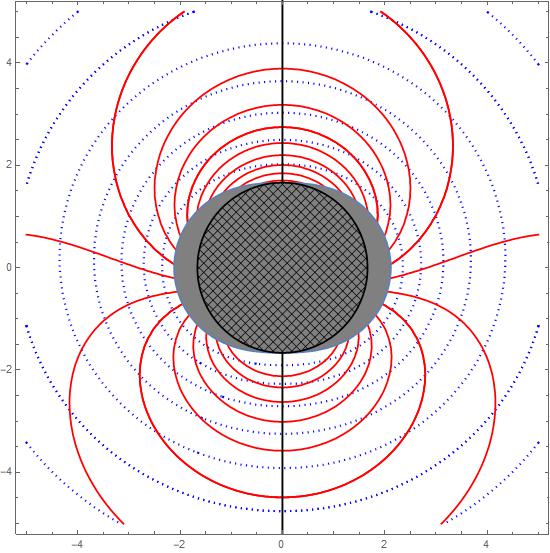}}
\subfigure[][$q$-metric with $q=0.60 M$, $a=0$, $\Omega = 0.116 M^{-1}$]{\includegraphics[width=0.26\textwidth]{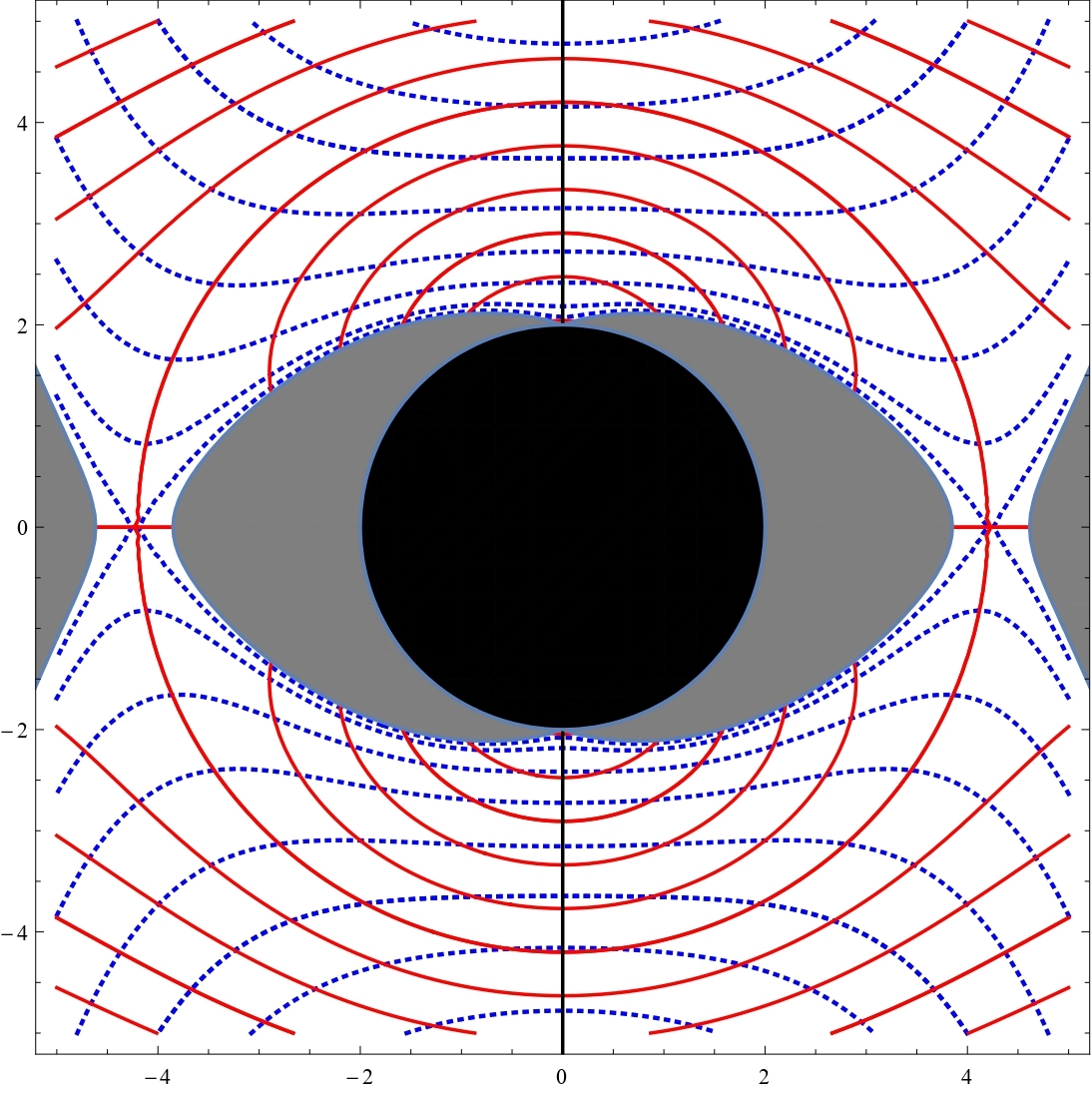}}
\subfigure[][$q$-metric with $q=0.60 M$, $a=0$, $\Omega = 0.118 M^{-1}$]{\includegraphics[width=0.26\textwidth]{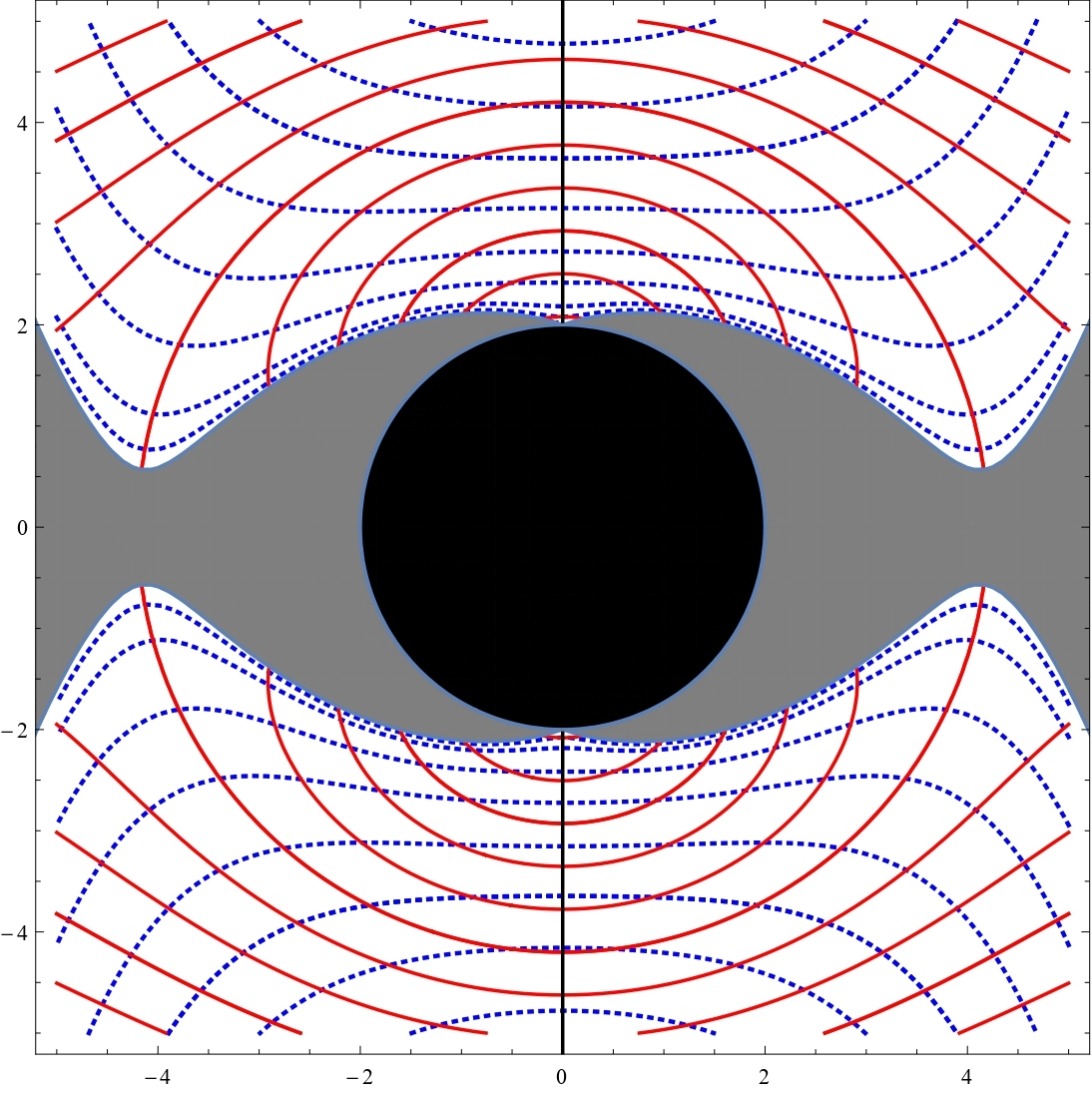}}
\subfigure[][$q$-metric with $q=1.00 M$, $a=0.90 M$, $\Omega = 0$]{\includegraphics[width=0.26\textwidth]{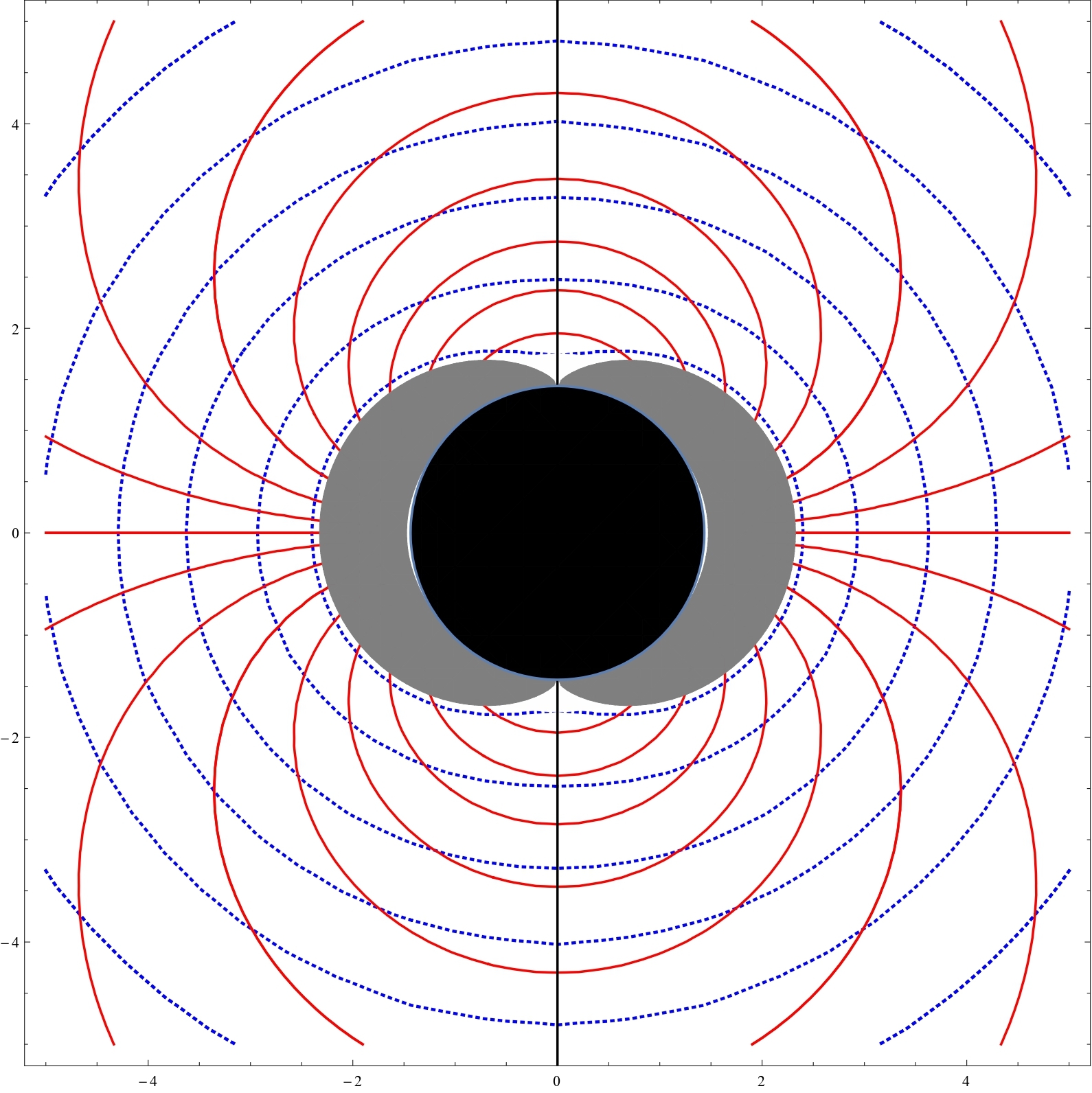}}
\subfigure[][$q$-metric with $q=-0.90 M$, $a=0.90 M$, $\Omega = 0$]{\includegraphics[width=0.26\textwidth]{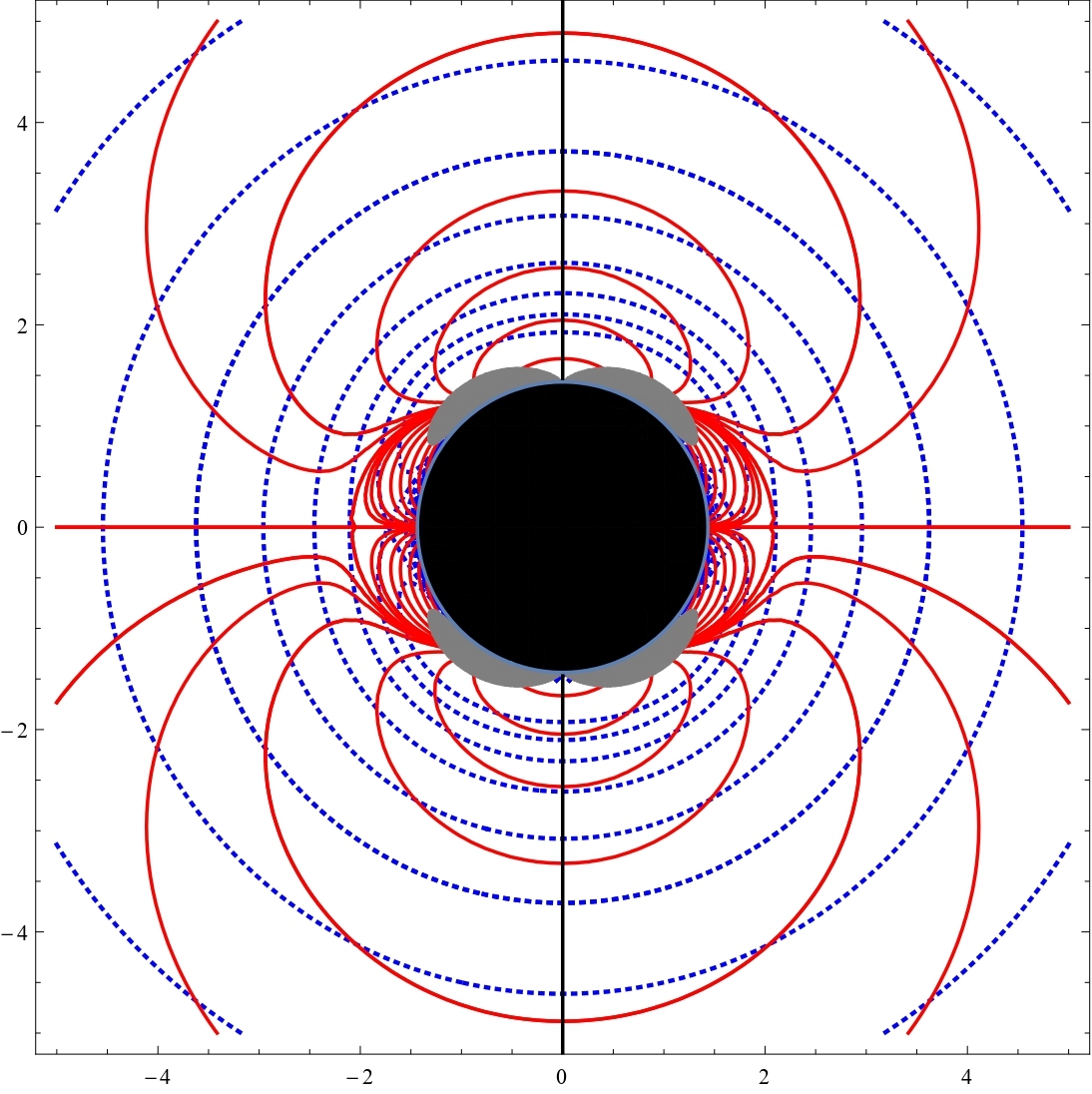}}
\caption{The gravitoelectric (blue dotted lines) and gravitomagnetic (red lines) potentials for the (Kerr-)NUT spacetime and for the $q$-metric. The crosshatched area is the region inside a black-hole horizon, the black area is the part bounded by a naked singularity, and the gray shaded area is the region where the potentials are not defined because the Killing vector field is spacelike. \label{fig:ex2}}
\end{figure*}

Fig.$\,$\ref{fig:ex2}$\,$(e) and (f) show the potentials for the nonspinning $q$-metric ($a = 0$) for two different values of $\Omega$.  Roughly speaking, the pictures look like squashed versions of the Schwarzschild case, which is of course an effect of the non-zero quadrupole moment.  Fig.$\,$\ref{fig:ex2}$\,$(g) and (h) show the potentials for the spinning $q$-metric ($a \neq 0$) where the observers are non-rotating ($\Omega = 0$).

\section{Discussion}

This paper is based on the observation that for a stationary solution to Einstein's vacuum field equation there are two scalar gravitational potentials. While the gravitoelectric potential is a general-relativistic generalization of the Newtonian potential the gravitomagnetic one has no non-relativistic analogue. There are many classical and quantum methods to operationally realize these potentials. The gravitoelectric potential can be interpreted as a height and the gravitomagnetic potential as a measure of the latitude. This means that these two potentials might be used as a physically given reference system in the vicinity of the rotating Earth. More precisely, if some degenerate cases are excluded they can be used as two of the three coordinates one needs for parametrizing 3-dimensional space. As the general-relativistic geoid is a particular level surface of the gravitoelectric potential, the intersection lines of the gravitomagnetic equipotential surfaces with the geoid give a latitudinal coordinatization of the geoid. 

Unfortunately, the gravitomagnetic effects on Earth are very small though they have been measured by LAGEOS and by Gravity Probe B via the Lense-Thirring and the Schiff effect. As we have seen, the Kerr parameter of the Earth also influences the gravitoelectric potential. However, its influence is just one order below the current accuracy of gravimeters. So, the next generation of instruments measuring the gravitoelectric and gravitomagnetic effects will be sensitive to the influence of the Earth's rotation on its gravitational field. The latter may also become observable with the help of the gravitomagnetic clock effect \cite{CohenMashhoon1993,Hackmann:2014aga}. 

The full gravitational field of the Earth can be given by a multipole expansion of the two potentials and the spatial metric $\gamma_{ij}$, see \eqref{metricdecomp}.  These components of the full metric can be measured with stationary and moving clocks, interferometers and gyroscopes. Within this framework, any adiabatic change of the gravitational field can be described through time-dependent multipole parameters. 

It has to be worked out how the potentials as well as the spatial metric can be measured from space, that is, with moving clocks or with GRACE-like constellations. A particular question is the following: Whereas for the gravitoelectric potential it is possible to measure potential differences, in particular in terms of the redshift of clocks, for the gravitomagnetic potential the measurement methods discussed above only yield the gradient. It would be interesting to find out if there is a method to measure differences of the gravitomagnetic potential. 

\section*{Acknowledgment}
We thank Eva Hackmann and Dennis Philipp for fruitful discussions. We acknowledge the support by the Deutsche Forschungsgemeinschaft (DFG, German Research Foundation) under Germany’s Excellence Strategy-EXC-2123 “QuantumFrontiers” -- Grant No. 390837967 and the CRC 1464 “Relativistic and Quantum-based Geodesy” (TerraQ).


%

\end{document}